\documentstyle[12pt]{article}
\date{\small}
\title{\bf On the Quantum Aspects of Geophysics}
\author{F. Darabi\thanks{e-mail:
f.darabi@azaruniv.edu}\\{\small Department of Physics, Azarbaijan
University of Tarbiat Moallem, 53714-161, Tabriz, Iran .} }
\begin{document}
\maketitle
\begin{abstract}
We introduce a simple quantum mechanical justification for the
formation of folded mountains. It is very appealing to develop
this idea to a theory of {\it Quantum Geophysics}
\end{abstract}

Quantum mechanics is the basis of our knowledge about the
microscopic world. The basic element in this theory is the wave
function $\Psi(\vec{r}, t)$ which, in principle, is assumed to
include all the physical information about a quantum mechanical
system, and its dynamics is governed by the Schr\"{o}dinger
equation. However, the domain to which one may apply the rules of
quantum mechanics is not limited to the microscopic world. Quantum
cosmology is an outstanding indication of this claim. In this
theory, the whole universe, as a single system, is described by a
wave function $\Psi(h_{ij}, \phi)$ which is believed to
incorporate all possible three geometries and matter field
configurations. This wave function satisfies a constraint called
Wheeler-DeWitt equation which plays the role of Schr\"{o}dinger
equation. One of the major tasks in quantum cosmology is to look
for a {\it good} correspondence between the quantum cosmological
predictions and the classical solutions for the geometry and
matter field configurations. This is called the classical limit of
quantum cosmology. In both quantum mechanics and quantum cosmology
the quantity $|\Psi|^2$ plays a key important role in the sense
that it provides us with the probability of finding a definite
configuration. Especially, in the quantum cosmology $|\Psi|^2$ may
be used to investigate the correspondence between the solutions of
Wheeler-DeWittt equation, in one hand, and the classical paths, on
the other hand. One is then tempted to try to test the
applicability of quantum mechanics to other macroscopic systems.
In this regard, one may pay attention to the Earth, as a single
system, and consider the formation of mountains on its crust.

Plate tectonics is a relatively new theory that has revolutionized
the way geologists think about the Earth. Since the mid-1960s, the
plate tectonic theory has been rigorously and successfully tested.
According to this theory, the surface of the Earth is broken into
a series of relatively thin, but large plates. The size and
position of these plates change over time. The fastest plates e.g.
Pacific Plates, move at over 10cm a year. The African plate moves
about 25 mm per year, whereas the Australian plate moves about 60
mm per year. The edges of these plates, where they move against
each other, are sites of intense geologic activities, such as
earthquakes, volcanoes, and mountain building. Folded mountains
are the most common type of mountain. They are created by tectonic
plates pushing against each other. This creates intense pressure.
Therefore, the only direction for these mountains to move is up.
The formation of folded mountains on Earth's surface can be
compared to the process of wrinkling a flexible rug on a floor. As
horizontal pressure is applied along the edge of a rug, folds
appear perpendicular to the direction of pressure. On Earth, as
horizontal pressure is applied to rocks, ridges and valleys form
perpendicular to the direction of the pressure.

According to quantum theory, every mass $m$, with momentum $p$ has
a wavelength $\lambda=\frac{h}{p}$, where $h$ is the planck
constant. In quantum cosmology the universe is simply assumed to
be a particle in a definite potential and one usually takes the
units so that $h=1$. This is because people are usually interested
in quantum cosmology to describe the birth of the universe and
find the correct boundary conditions to investigate properly the
singularity problems in general relativity, and the value of $h$
is not of prime importance. In other words, to the extent we are
dealing with the characteristic behavior of the wave function of
the universe, namely $|\Psi|^2$, we may ignore $h$.

It is therefore appealing to assume a relation $\lambda \sim
p^{-1}$ to describe the formation of folded mountains on the basis
of {\it wave-particle} interpretation of quantum mechanics.
Suppose a particle with total energy $E=mgy_0$ is subjected to the
following one-dimensional gravitational potential
$$
V(x)=\left \{ \begin{array}{ll}
\infty \hspace{11mm} x\leq 0\\
\\
a x \hspace{11mm} x>0
\end{array}\right.
$$
where $a=\frac{E}{L}$ is a constant and $L$ is a length over which
the particle is limited to its classical motion. Without solving
the Schr\"{o}dinger equation one may obtain the characteristic
features of the particle's wave function. At $x=0$, the wave
function is zero due to the infinite potential. At $x\approx 0$,
the wavelength is small due to the high velocity of the particle
and the amplitude is small, as well, due to the fact that the
probability of observing this fast particle at $x\approx 0$ is
small. As $x$ becomes larger the wavelength and the amplitude of
the particle's wave gets larger, as well, due to the fact that the
velocity of the particle decreases with $x$. At $x=L$, the wave
has its maximum wavelength and amplitude, because the region $x>L$
is classically forbidden for the particle, so the associated wave
will lose the oscillating behavior and the amplitude starts
decreasing. If we now evaluate $\Psi(x)$ and plot $|\Psi(x)|^2$ in
the interval $0\leq x \leq L$, as in Fig.1, we then surprisingly
obtain almost the same general pattern of a folded mountain as
viewed from its transverse cross section, starting from its flat
point and ending at the top. We believe this similarity is not a
coincidence. One may think that when two plates collide each other
the velocity of their front boundaries decreases considerably
whereas the whole plates keep their motion, as before. Therefore,
as we move from inside the plates towards colliding boundaries the
velocity of interior boundary layers decreases smoothly down to
the velocity of the front boundaries. We now associate a
wavelength to each layer which may be interpreted as a particle
with varying velocity in the above potential.  The wavelength of
the interior boundary layers will be shorter and that of front
boundaries will be longer due to the faster and slower motions of
these layers, respectively. On the other hand, the amplitude
$|\Psi|$ for the interior boundary layers are smaller and that of
front boundaries are larger due to the same reason. We may now
interpret the whole interior part of each plate as an infinite
potential and the boundary layers as a particle, with varying
velocity, obeying the above linear potential. Each front boundary
layer then represents a particle with total energy $E=mgy_0$ at a
distance $L$ from the infinite potential. The associated wave of
this layer will have the longest wavelength and the biggest
amplitude; see the last wavelength at the right side in Fig.1, for
a typical value $L=200$. Combining these behaviors of two
colliding front boundaries and taking into account $|\Psi|^2$
leads to a pattern which resembles the central wide and top
portion of the folded mountain, see the central portion in Fig.2.

On the contrary, the last interior boundary layer of each plate
represents a particle with total energy $E=\frac{p^2}{2m}$
approaching the infinite potential. The associated wave of these
layers will have the shortest wavelength and vanishing amplitude;
see the first wavelength at the left side in Fig.1. One may
correspond this behavior, by taking $|\Psi|^2$, to the flat sides
of the folded mountain, see Fig.2. In general, if we associate
these wave properties, to each two-dimensional plate, as $\Psi_1$
and $\Psi_2$, then the folded mountains will be nothing but the
manifestation of a wave interference phenomena, namely
$|\Psi_1+\Psi_2|^2$, which resembles the general pattern of these
mountains. In this way, one may correspond $|\Psi|^2$ to the
height of the mountain at a given point, provided we interpret
$|\Psi|^2$ as the probability of finding the geologic matter at
that point. The top point of a mountain is then a point for which
$|\Psi|^2$ is maximum, namely, the probability of matter
accumulation at this point is maximum.

In this regard, one may assume that all types of mountains, hills
and valleys are produced due to the very complicate wave
interference phenomena originating from the interior dynamics of
hot mantle. This is an interesting realization of {\it
wave-particle} duality at this large scale phenomena.

It seems, as in quantum mechanics ( or quantum cosmology ), that
one may go beyond and look for a Schr\"{o}dinger-like equation, in
Geophysics, whose solution in principle may have complete
information about the formation of all types of mountains, hills
and valleys. If we define the state ket $|\alpha>$ to represent
the whole crust of the Earth, one may then write
$$
<\alpha|\alpha>=\int d^2X |<X|\alpha>|^2=1,
$$
where $|<X|\alpha>|^2$ determines the shape of the crust at
position $X$, on the two dimensional surface of the Earth.

\newpage
{\large {\bf Figure captions}}
\\
\\
Fig. 1. $|\Psi|^2$ for a particle in the potential $V(x), 0\leq x \leq L$. \\
\\
Fig. 2. $|\Psi|^2$ for two particles in the piecewise potentials
$V(x), 0\leq x \leq L$ and $V(2L-x), L\leq x \leq 2L$.
\end{document}